\documentclass[11pt]{article}
\usepackage[top=1.0in,right=1.0in,left=1.0in,bottom=1.75in]{geometry}
\usepackage{latexsym}
\usepackage{amssymb}
\usepackage{amsmath,amsfonts,amssymb,amsthm}
\usepackage{float}
\usepackage{graphics,graphicx}
\usepackage{tabularx}
\usepackage{amsfonts}
\usepackage{amsmath}
\usepackage{enumerate}
\usepackage{booktabs}
\usepackage{multirow}
\usepackage{sectsty}
\usepackage[affil-it,auth-sc]{authblk}
\usepackage{subcaption}
\usepackage[usenames,dvipsnames]{xcolor}

\usepackage{tabu}

\usepackage{fancyhdr}            % NB: E=pagina pari, O=pagina dispari, H=intestazione, F=piede pagina, L,R,C=sinistra,destra,centro ; \leftmark = capitolo ; \rightmark = paragrafo
\fancypagestyle{plain}{%
  \fancyhead{}                                     % Pulisce il campo della testata
  \fancyfoot[CE,CO]{\thepage}       % Numero pagina in basso ed esterno in grassetto
  \fancyhead[CE,CO]{\textcolor[rgb]{0.64,0.15,0.15}{Published in \emph{Proceedings of the Royal Society A} (2014) \textbf{470}: 20140232
\\
doi: http://dx.doi.org/10.1098/rspa.2014.0232}}
\setlength{\headheight}{14.5pt}}
\newcommand{\footmsg}[1]{%
  \let\temp\thempfn%
  \def\thempfs{}
  \footnotetext{#1}
  \let\tempfn\temp}

\begin{document}

%Definitions: general
\newcommand{\singlespace} {\baselineskip=12pt
\lineskiplimit=0pt \lineskip=0pt }
\def\ds{\displaystyle}

%Definitions: equations
\newcommand{\beq}{\begin{equation}}
\newcommand{\eeq}{\end{equation}}
\newcommand{\lb}{\label}
\newcommand{\beqar}{\begin{eqnarray}}
\newcommand{\eeqar}{\end{eqnarray}}
\newcommand{\barr}{\begin{array}}
\newcommand{\earr}{\end{array}}

\newcommand{\jump}{\parallel}

\def\c{{\circ}}

\newcommand{\Ehat}{\hat{E}}
\newcommand{\That}{\hat{\bf T}}
\newcommand{\Ahat}{\hat{A}}
\newcommand{\chat}{\hat{c}}
\newcommand{\shat}{\hat{s}}
\newcommand{\khat}{\hat{k}}
\newcommand{\muhat}{\hat{\mu}}
\newcommand{\mc}{M^{\scriptscriptstyle C}}
\newcommand{\mei}{M^{\scriptscriptstyle M,EI}}
\newcommand{\mec}{M^{\scriptscriptstyle M,EC}}

\newcommand{\hbeta}{{\hat{\beta}}}
\newcommand{\rec}[2]{\left( #1 #2 \ds{\frac{1}{#1}}\right)}
\newcommand{\rep}[2]{\left( {#1}^2 #2 \ds{\frac{1}{{#1}^2}}\right)}
\newcommand{\derp}[2]{\ds{\frac {\partial #1}{\partial #2}}}
\newcommand{\derpn}[3]{\ds{\frac {\partial^{#3}#1}{\partial #2^{#3}}}}
\newcommand{\dert}[2]{\ds{\frac {d #1}{d #2}}}
\newcommand{\dertn}[3]{\ds{\frac {d^{#3} #1}{d #2^{#3}}}}

\def\bob{{\, \underline{\overline{\otimes}} \,}}

\def\ob{{\, \underline{\otimes} \,}}
\def\scalp{\mbox{\boldmath$\, \cdot \, $}}
\def\gdp{\makebox{\raisebox{-.215ex}{$\Box$}\hspace{-.778em}$\times$}}

\def\daa{\makebox{\raisebox{-.050ex}{$-$}\hspace{-.550em}$: ~$}}

\def\mK{\mbox{${\mathcal{K}}$}}
\def\cK{\mbox{${\mathbb {K}}$}}

%Definitions: integrals
\def\Xint#1{\mathchoice
   {\XXint\displaystyle\textstyle{#1}}%
   {\XXint\textstyle\scriptstyle{#1}}%
   {\XXint\scriptstyle\scriptscriptstyle{#1}}%
   {\XXint\scriptscriptstyle\scriptscriptstyle{#1}}%
   \!\int}
\def\XXint#1#2#3{{\setbox0=\hbox{$#1{#2#3}{\int}$}
     \vcenter{\hbox{$#2#3$}}\kern-.5\wd0}}
\def\ddashint{\Xint=}
\def\fpint{\Xint=}
\def\dashint{\Xint-}
\def\cpvint{\Xint-}
\def\intl{\int\limits}
\def\cpvintl{\cpvint\limits}
\def\fpintl{\fpint\limits}
\def\ointl{\oint\limits}

\def\half{{\scriptstyle{\frac{1}{2}}}}

\def\bA{{\bf A}}
\def\ba{{\bf a}}
\def\bB{{\bf B}}
\def\bb{{\bf b}}
\def\bc{{\bf c}}
\def\bC{{\bf C}}
\def\bD{{\bf D}}
\def\bE{{\bf E}}
\def\be{{\bf e}}
\def\bbf{{\bf f}}
\def\bF{{\bf F}}
\def\bG{{\bf G}}
\def\bg{{\bf g}}
\def\bi{{\bf i}}
\def\bH{{\bf H}}
\def\bK{{\bf K}}
\def\bL{{\bf L}}
\def\bM{{\bf M}}
\def\bN{{\bf N}}
\def\bn{{\bf n}}
\def\bm{{\bf m}}
\def\b0{{\bf 0}}
\def\bo{{\bf o}}
\def\bX{{\bf X}}
\def\bx{{\bf x}}
\def\bP{{\bf P}}
\def\bp{{\bf p}}
\def\bQ{{\bf Q}}
\def\bq{{\bf q}}
\def\bR{{\bf R}}
\def\bS{{\bf S}}
\def\bs{{\bf s}}
\def\bT{{\bf T}}
\def\bt{{\bf t}}
\def\bU{{\bf U}}
\def\bu{{\bf u}}
\def\bv{{\bf v}}
\def\bw{{\bf w}}
\def\bW{{\bf W}}
\def\by{{\bf y}}
\def\bz{{\bf z}}

\def\T{{\bf T}}
\def\Te{\textrm{T}}

\def\e{{\rm{e}}}
\def\Id{{\bf I}}
\def\p{{\rm{p}}}
\def\t{{\rm{t}}}
\def\bxi{\mbox{\boldmath${\xi}$}}
\def\balpha{\mbox{\boldmath${\alpha}$}}
\def\bbeta{\mbox{\boldmath${\beta}$}}
\def\bepsilon{\mbox{\boldmath${\epsilon}$}}
\def\bvarepsilon{\mbox{\boldmath${\varepsilon}$}}
\def\bomega{\mbox{\boldmath${\omega}$}}
\def\bphi{\mbox{\boldmath${\phi}$}}
\def\bsigma{\mbox{\boldmath${\sigma}$}}
\def\bfeta{\mbox{\boldmath${\eta}$}}
\def\bDelta{\mbox{\boldmath${\Delta}$}}
\def\btau{\mbox{\boldmath $\tau$}}

\def\tr{{\rm tr}}
\def\dev{{\rm dev}}
\def\div{{\rm div}}
\def\Div{{\rm Div}}
\def\Grad{{\rm Grad}}
\def\grad{{\rm grad}}
\def\Lin{{\rm Lin}}
\def\Sym{{\rm Sym}}
\def\Skw{{\rm Skew}}
\def\abs{{\rm abs}}
\def\Re{{\rm Re}}
\def\Im{{\rm Im}}
\def\sign{{\rm sign}}

\def\capB{\mbox{\boldmath${\mathsf B}$}}
\def\capC{\mbox{\boldmath${\mathsf C}$}}
\def\capD{\mbox{\boldmath${\mathsf D}$}}
\def\capE{\mbox{\boldmath${\mathsf E}$}}
\def\capG{\mbox{\boldmath${\mathsf G}$}}
\def\tcapG{\tilde{\capG}}
\def\capH{\mbox{\boldmath${\mathsf H}$}}
\def\capK{\mbox{\boldmath${\mathsf K}$}}
\def\capL{\mbox{\boldmath${\mathsf L}$}}
\def\capM{\mbox{\boldmath${\mathsf M}$}}
\def\capR{\mbox{\boldmath${\mathsf R}$}}
\def\capW{\mbox{\boldmath${\mathsf W}$}}

%imaginary unit
\def\i{\mbox{${\mathrm i}$}}

\def\mC{\mbox{\boldmath${\mathcal C}$}}

\def\mB{\mbox{${\mathcal B}$}}
\def\mE{\mbox{${\mathcal{E}}$}}
\def\mL{\mbox{${\mathcal{L}}$}}
\def\mK{\mbox{${\mathcal{K}}$}}
\def\mV{\mbox{${\mathcal{V}}$}}

\def\C{\mbox{\boldmath${\mathcal C}$}}
\def\E{\mbox{\boldmath${\mathcal E}$}}

%Definitions: journals
\def\ACME{{ Arch. Comput. Meth. Engng.\ }}
\def\ARMA{{ Arch. Rat. Mech. Analysis\ }}
\def\AMR{{ Appl. Mech. Rev.\ }}
\def\ASCEEM{{ ASCE J. Eng. Mech.\ }}
\def\acta{{ Acta Mater. \ }}
\def\AMM{{ Acta Metall. Mater. \ }}
\def\CMAME {{ Comput. Meth. Appl. Mech. Engrg.\ }}
\def\CRAS{{ C. R. Acad. Sci., Paris\ }}
\def\EFM{{ Eng. Fracture Mechanics\ }}
\def\EJMA{{ Eur.~J.~Mechanics-A/Solids\ }}
\def\IJES{{ Int. J. Eng. Sci.\ }}
\def\IJF{{\it Int. J. Fracture\ }}
\def\IJMS{{ Int. J. Mech. Sci.\ }}
\def\IJNAMG{{ Int. J. Numer. Anal. Meth. Geomech.\ }}
\def\IJP{{ Int. J. Plasticity\ }}
\def\IJSS{{ Int. J. Solids Structures\ }}
\def\IngA{{ Ing. Archiv\ }}
\def\JAM{{ J. Appl. Mech.\ }}
\def\JAP{{ J. Appl. Phys.\ }}
\def\JEM{{J. Engrg. Mech., ASCE\ }}
\def\JE{{ J. Elasticity\ }}
\def\JM{{ J. de M\'ecanique\ }}
\def\JMPS{{ J. Mech. Phys. Solids\ }}
\def\Macro{{ Macromolecules\ }}
\def\MOM{{ Mech. Materials\ }}
\def\MMS{{ Math. Mech. Solids\ }}
\def\MMT{{\it Metall. Mater. Trans. A}}
\def\MPCPS{{ Math. Proc. Camb. Phil. Soc.\ }}
\def\MSE{{ Mater. Sci. Eng.}}
\def\PMPS{{ Proc. Math. Phys. Soc.\ }}
\def\PRE{{ Phys. Rev. E\ }}
\def\PRSL{{ Proc. R. Soc.\ }}
\def\rock{{ Rock Mech. and Rock Eng.\ }}
\def\QAM{{ Quart. Appl. Math.\ }}
\def\QJMAM{{ Quart. J. Mech. Appl. Math.\ }}
\def\SCRMAT{{ Scripta Mater.\ }}
\def\SM{{\it Scripta Metall. }}

% segue comando pazzesco di zaccaria

\def\salto#1#2{
%\left[\mbox{\hspace{-#1em}}\left[#2\right]\mbox{\hspace{-#1em}}\right]}
\left[\mbox{\hspace{-#1em}}\left[#2\right]\mbox{\hspace{-#1em}}\right]}

\def\medio#1#2{
%\left[\mbox{\hspace{-#1em}}\left[#2\right]\mbox{\hspace{-#1em}}\right]}
\mbox{\hspace{-#1em}}<#2>\mbox{\hspace{-#1em}}}

%\def\salto{
%\left[\mbox{\hspace{-#1em}}\left[#2\right]\mbox{\hspace{-#1em}}\right]}
%[[#1]]}

%\newcommand{\salto}[1]{\ds{\|#1\|}}

%dopodiche' scrivi il comando
%\salto{*}{**}
%dove * e' un numero mentre ** e' quello che tu vuoi tra parentesi.
%Il numero serve a far si che le due parentesi quadre che comporranno
%l'unica parentesi che tu vuoi siano ben posizionate. Tale numero varia a
%seconda di chi sia **.

\title{An Elastica Arm Scale}

\author{F. Bosi, D. Misseroni, F. Dal Corso and D. Bigoni$^{(0)}$\\
\normalsize{federico.bosi@unitn.it; diego.misseroni@unitn.it; francesco.dalcorso@unitn.it}
\\
\small{DICAM, University of Trento, via Mesiano 77, I-38123 Trento, Italy}\\ 
\small{(0) Corresponding author bigoni@ing.unitn.it
}}
%\date{}

\maketitle

\begin{abstract}
The concept of \lq deformable arm scale' (completely different from a traditional rigid arm balance) is theoretically
introduced and experimentally validated. The idea is not intuitive, but is the result of nonlinear equilibrium kinematics of rods inducing configurational forces, so that
deflection of the arms becomes necessary for the equilibrium, which would
be impossible for a rigid system. In particular, the rigid arms of usual scales are replaced by a flexible elastic lamina,
free of sliding in a frictionless and inclined sliding sleeve, which can reach a unique equilibrium configuration when two vertical dead loads are applied.
Prototypes realized to demonstrate the feasibility of the system show a high accuracy in the measure of load within a certain range of use.
It is finally shown that the presented results are strongly related to snaking of confined beams, with implications on locomotion of serpents, plumbing, and smart oil drilling.
\end{abstract}


\vspace{10 mm}

\setcounter{equation}{0}

\section{Introduction}

For millennia the (equal and unequal) arm balance scales have been used, and still are used (see the overview \cite{librobilance}), to measure weight
by exploiting equilibrium of a rigid lever, so that a deformation of the arms would merely represent an undesired effect.
On the other hand, equilibrium is always satisfied for a spring balance, where the weight measure is directly linked to deformation and a counterweight
is not needed.
A new paradigm, based on exploitation of nonlinear kinematics and configurational mechanics of elastic rods, is proposed here for a scale with deformable arms, where an inflected equilibrium
configuration can be exploited to measure weight. In a sense, the proposed balance is a sort of
combination between a rigid arm and a spring balance, because equilibrium and deformation are both simultaneously exploited. Therefore,
the concept introduced here differs completely from that underlying traditional scale design, so that the proposed device can work with or without
a counterweight.

\begin{figure}[!htcb]
\begin{center}
\includegraphics[width= 14 cm]{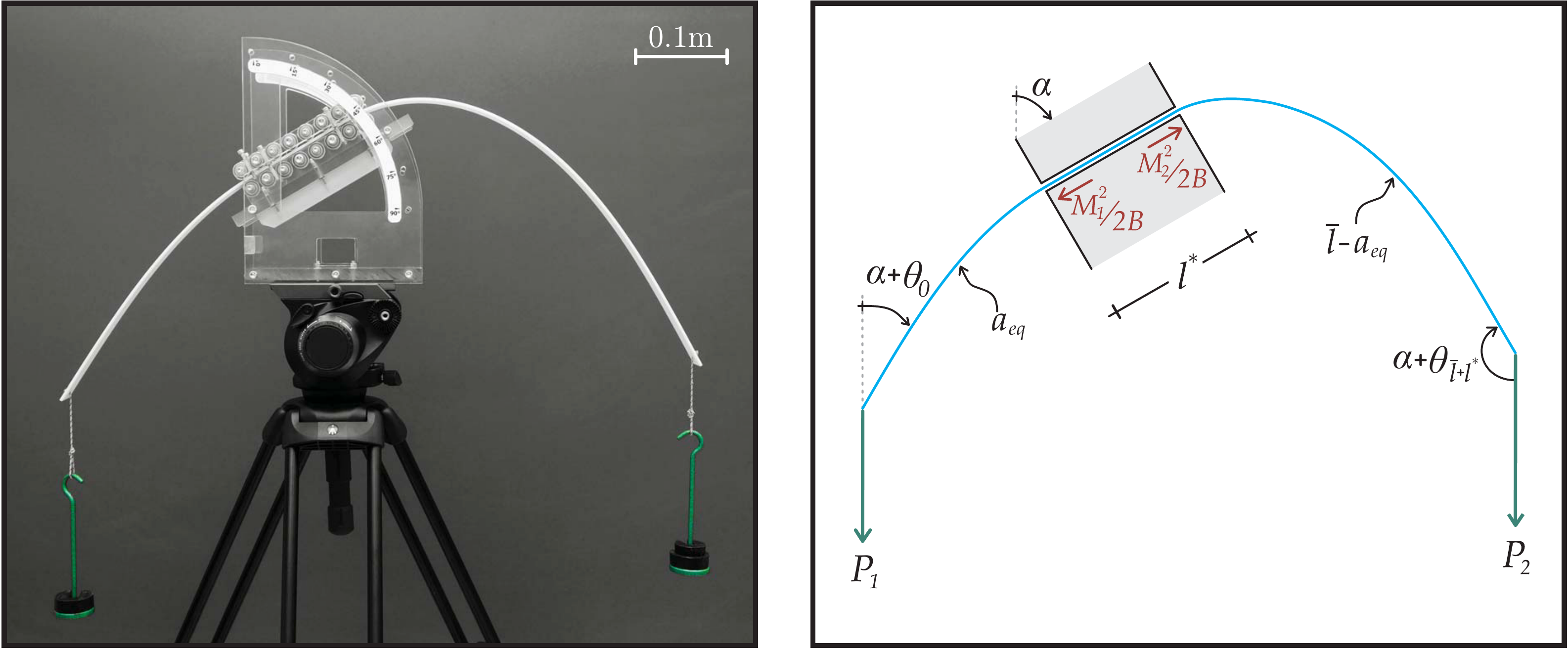}
\caption{(Left) Prototype 1 and (right) scheme of the deformable arm scale.
The rod used in the prototype is made up of a solid polycarbonate elastic lamina of bending stiffness $B$=0.20 Nm$^2$ and total
length $\bar{l}+ l^*$=0.98 m with ends subject to dead loads $P_1$=2.03 N and $ P_2 $=2.52 N. The lamina can slide into a frictionless
sliding sleeve (realized with 8 roller pairs) of length $ l^*$=0.148 m,
inclined at an angle $\alpha$=60$^\circ$ with respect to the vertical direction.
The theoretical value of the length defining the equilibrium configuration is $a_{eq}$=0.35 m,
while the value measured on the prototype is equal to 0.34 m.}
\label{system}
\end{center}
\end{figure}

The  \lq elastically
deformable arm scale' is shown on the left of Fig. \ref{system} (photo of prototype 1) as a realization of the scheme reported on the right of Fig. \ref{system},
where an elastic rod (inclined at an angle $\alpha\in[0,\pi/2]$
with respect to the two vertical dead loads
applied at its edges) is free of sliding in a frictionless sleeve of length $l^*$. For given loads ($P_1$ and $P_2$), the scale admits an equilibrium configuration,
possible by virtue of the flexural deformation of the arms
(would these be rigid,  the equilibrium would be trivially violated).
%%%%%%%%%%%%%%%%%%%%%%%%%%%%%%%%%%%%%%%%%%%%%%%%%%%%%%%%%%%%%%%%%%%%%%
%%%%%%%%%%%%%%%%%%%%%%%%%%%%%%%%%%%%%%%%%%%%%%%%%%%%%%%%%%%%%%%%%%%%%%
This equilibrium configuration is inherently nonlinear, as it necessarily involves the presence of configurational or \lq Eshelby-like' forces \cite{bigonialtri},
but can be derived from stationarity of the total potential energy in a form suitable for direct calculations.
Therefore, the nonlinear equilibrium equations (section \ref{stocaz}) can be exploited to determine a
load from the measure of a configurational parameter (the length $a_{eq}$).
Considerations on the second variation of the total potential energy (section \ref{socmel}), show  that
the equilibrium configurations of the scale are unstable, a feature that may enhance the precision of the
load measure and that does not prevent the feasibility of the
scale, as shown through realization of two \lq proof-of-concept prototypes' (section \ref{vankuelen}).
Furthermore, a sensitivity analysis and the experiments performed on the prototypes
indicate that the deformable arm balance works correctly and that can be more performing than traditional balances in certain load ranges.

It can be finally mentioned that the elastic lamina in the sliding sleeve (realizing the deformable arm scale) can be viewed as a \lq snake in a frictionless and tight channel', so that our results  demonstrate that Eshelby-like forces play a fundamental role in the problem of snaking of a confined elastic rod, with implications on smart oil drilling \cite{beck}, plumbing,
and reptiles and fish locomotion \cite{gray1, gray2, gray3}.

\section{Flexural equilibrium through Eshelby-like forces} \lb{stocaz}

The system shown in Fig. \ref{system} (right) attains equilibrium because two forces exist, tangential to the sliding sleeve,
which can be interpreted as \lq configurational' (or \lq Eshelby-like' \cite{bigonialtri}), in the sense that they depend on
the configuration assumed by the system at equilibrium.
These forces and the equilibrium conditions of the system can be obtained for an inextensible elastic lamina of bending stiffness $B$
and total length $\bar{l}+l^{*}$  from  the first variation of the total potential energy of the system \cite{bigoni}
\begin{equation}
\begin{split}
\mathcal{V}(\theta(s),a)&=\intop_{0}^{a}B
\frac{\left[\theta^{'}\left(s\right)\right]^{2}}{2}ds+\intop_{a+l^{*}}^{\bar{l}+l^{*}}B
\frac{\left[\theta^{'}\left(s\right)\right]^{2}}{2}ds-
P_1\left[\cos\alpha \intop_{0}^{a}\cos\theta(s)ds-\sin\alpha\intop_{0}^{a}\sin\theta(s)ds \right]\\
&\ -P_2\left[-\cos\alpha \intop_{a+l^{*}}^{\bar{l}+l^{*}}\cos\theta(s)ds+
\sin\alpha\intop_{a+l^{*}}^{\bar{l}+l^{*}}\sin\theta(s)ds \right],
\end{split}
\label{EPT}
\end{equation}
where $s \in \left[0;\bar{l}+l^{*}\right]$ is a curvilinear coordinate, $\theta (s)$ is the rotation of the rod's axis, $a$
and $a+l^{*}$, are the curvilinear coordinates at which, respectively, the left arm terminates and the right one
initiates, so that  $\theta(s)=0$ for $s \in \left[a;a+l^{*}\right]$.
The parameter $a$, defining the position of the rod with
respect to the sliding sleeve, is variable, to be adjusted until the equilibrium configuration is reached.

Considering a small parameter $\epsilon$ and taking variations (subscript \lq \emph{var}') of an equilibrium configuration (subscript \lq \emph{eq}')
in the form $\theta=\theta_{eq}(s)+\epsilon\theta_{var}(s)$ and $a=a_{eq}+\epsilon a_{var}$,
four compatibility equations are obtained from a Taylor series expansion of the rotation field $\theta (\hat{s})$
for $\hat{s} = a$ and $\hat{s} = a+l^{*}$, namely (see \cite{tarzan} for details),
\beq\label{compatire}
\begin{array}{lll}
&\theta_{var}(a_{eq})=-a_{var} \theta_{eq}^{'}(a_{eq}),\qquad
&\theta_{var}(a_{eq}+l^*)=-a_{var} \theta_{eq}^{'}(a_{eq}+l^*),\\[3 mm]
&\ds\theta_{var}'(a_{eq})=-\frac{1}{2} a_{var} \theta_{eq}^{''}(a_{eq}),
&\ds \theta_{var}'(a_{eq}+l^*)=-\frac{1}{2} a_{var} \theta_{eq}^{''}(a_{eq}+l^*).
\end{array}
\eeq
Through integration by parts and consideration of the first two compatibility conditions (\ref{compatire}) and the static conditions at the free edges, $\theta_{eq}^{'}(0)=\theta_{eq}^{'}(\bar{l}+l^*)=0$,
the first variation of the total potential energy (\ref{EPT}) can be obtained  as
\begin{equation}
\begin{split}
\ds \delta_\epsilon\mathcal{V}&=-\intop_{0}^{a_{eq}}\left[B
\theta_{eq}^{''}-P_1\bigl(\cos\alpha \sin\theta_{eq}(s)+\sin\alpha \cos\theta_{eq}(s)\bigr)\right] \theta_{var}(s)ds\\
&\ -\intop_{a_{eq}+l^{*}}^{\bar{l}+l^{*}}\left[B
\theta_{eq}^{''}+P_2\bigl(\cos\alpha \sin\theta_{eq}(s)+\sin\alpha \cos\theta_{eq}(s)\bigr)\right] \theta_{var}(s)ds\\
&\
+\left\{\frac{B}{2}\Bigl[\theta_{eq}^{'}(a_{eq}+l^{*})^{2}-\theta_{eq}^{'}(a_{eq})^{2}\Bigr]-(P_1+P_2)\cos\alpha\right\}a_{var}
, \label{varprimaEPT}
\end{split}
\end{equation}
and imposed to vanish
(for every variation in the rotation field $\theta_{var}(s)$
and in the length $a_{var}$) to obtain the equilibrium configuration.  This is
governed by:

(i.) the elastica \cite{love} for the two arms of the lamina
\begin{equation}
B \theta_{eq}^{''}(s)-P_j\sin\left[\theta_{eq}(s) - (-1)^j \alpha\right]=0,
\label{eq:elasticaeq}
\end{equation}
where $j=1$ for the left arm ($s \in \left[0,a_{eq}\right]$) and $j=2$ for the right one ($s \in \left[a_{eq}+l^{*},\bar{l}+l^{*}\right]$),
and

(ii.) the rigid-body equilibrium condition along the sliding direction of the sleeve
\begin{equation}
(P_1+P_2)\cos\alpha+\underbrace{\frac{M_1^2-M_2^2}{2B}}_{\mbox{\small{Eshelby-like
Forces}}} = 0,
\label{eq:axialeq}
\end{equation}
where $M_1= B\theta_{eq}^{'}(a_{eq})$ and $M_2= B\theta_{eq}^{'}(a_{eq}+l^{*})$.

Condition (\ref{eq:axialeq})  reveals the presence of two so-called \lq Eshelby-like forces' \cite{bigonialtri},
 provided by the sliding sleeve at its left and right ends and generated by the flexural deformation of the left and right arms, respectively,
 which define the equilibrium condition of the system
 and are the key concept of the deformable arm scale.

The rotations at the free ends at equilibrium, $\theta_{0}=\theta_{eq}(0)$ and $\theta_{\bar{l}+l^{*}}=\theta_{eq}(\bar{l}+l^{*})$,
can be obtained
by double integration of the elastica (\ref{eq:elasticaeq}), leading to the following conditions
\begin{equation}
 \ds a_{eq} \sqrt{\frac{P_1}{B}} = \mathcal{K}\left(\kappa_1\right)-\mathcal{K}\left(m_1,\kappa_1\right),
\qquad
\ds \left(\bar{l}-a_{eq}\right) \sqrt{\frac{P_2}{B}}  = \mathcal{K}\left(\kappa_2\right)-\mathcal{K}\left(m_2,\kappa_2\right),
\label{tratto1}
\end{equation}
where $\mathcal{K}\left(\kappa_j\right)$ and $\mathcal{K}\left(m_j,\kappa_j\right)$ are respectively the complete and incomplete elliptic integral
of the first kind
\begin{equation}
\mathcal{K}(\kappa_j)=\intop_{0}^{\frac{\pi}{2}}\frac{d\phi_j}{\sqrt{1-\kappa_j^{2}\sin^{2}\phi_j}},
\qquad \mathcal{K}(m_j,\kappa_j)=\intop_{0}^{m_j}\frac{d\phi_j}{\sqrt{1-\kappa_j^{2}\sin^{2}\phi_j}}, \qquad j=1,2.
\end{equation}
and
\beq
\begin{array}{lllllll}
\ds
&\ds\kappa_1=\sin\frac{\theta_{0}+\alpha+\pi}{2},
& \ds m_1=\arcsin\left[\frac{\sin\frac{\alpha+\pi}{2}}{\kappa_1}\right],
&\ds \kappa_1\sin\phi_1(s)=\sin\frac{\theta_{eq}(s)+\alpha+\pi}{2},
\\[5mm]
&\ds \kappa_2=\sin\frac{\theta_{\bar{l}+l^{*}}+\alpha}{2},
&\ds m_2=\arcsin\left[\frac{\sin\frac{\alpha}{2}}{\kappa_2}\right],
&\ds \kappa_2\sin\phi_2(s)=\sin\frac{\theta_{eq}(s)+\alpha}{2}.
\earr
\eeq
Further integration of the elastica (\ref{eq:elasticaeq}) leads to the
solution for the rotation field at equilibrium
\begin{equation}\lb{rotation1}
\theta_{eq}(s)=\left\{
\barr{lll}
\pi- 2\arcsin\left[\kappa_1 \textup{sn}\left(\mathcal{K}(\kappa_1)-\ds\sqrt{\frac{P_1}{B}} s,\kappa_1\right)\right]-\alpha,
&\,\, s\in\left[0,a_{eq}\right],\\[6mm]
2\arcsin\left[\kappa_2 \textup{sn}\left(\ds\sqrt{\frac{P_2}{B}}(s-a_{eq}-l^{*})+\mathcal{K}(m_2,\kappa_2),\kappa_2\right)\right]-\alpha,
&\,\, s\in\left[a_{eq}+l^{*},\bar{l}+l^{*}\right]
\earr
\right.
\end{equation}
where sn is the Jacobi sine amplitude function.
Since the solution (\ref{rotation1}) implies
\beq
B \theta_{eq}^{'}(a_{eq})^{2}=2 P_1 \left[\cos(\theta_0+\alpha)- \cos\alpha\right],\qquad
B \theta_{eq}^{'}(a_{eq}+l^{*})^{2}=2 P_2 \left[\cos\alpha- \cos(\theta_{\bar{l}+l^{*}}+\alpha)\right],
\eeq
the equilibrium along the sliding direction of the sleeve (\ref{eq:axialeq}) can be expressed as a \lq geometrical condition' of equilibrium,
which relates the angles at the free edges to the two applied vertical dead loads as
\beq
\lb{evviva}
P_1 \cos(\alpha+\theta_0)+P_2\cos(\alpha+\theta_{\bar{l}+l^{*}})=0,
\eeq
and represents the balance of axial thrust of the deformable scale
($0\leq\alpha+\theta_0\leq\alpha$ and $\pi/2\leq\alpha+\theta_{\bar{l}+l^{*}}\leq\pi$).

When $\alpha+\theta_{\bar{l}+l^{*}} = \pi/2$ the equilibrium eq. (\ref{evviva}) implies $P_1=0$, so that {\it a counterweight
is not needed}, Fig. \ref{fiore3}.
%%%%%%%%%%%%%%%%%%%%%%%%%%%%%%%%%%%%%%%%%%%%%%%%%%%%%%%%%%%%%%%%%%%%%%
\begin{figure}[!htcb]
\begin{center}
\includegraphics[width= 8 cm]{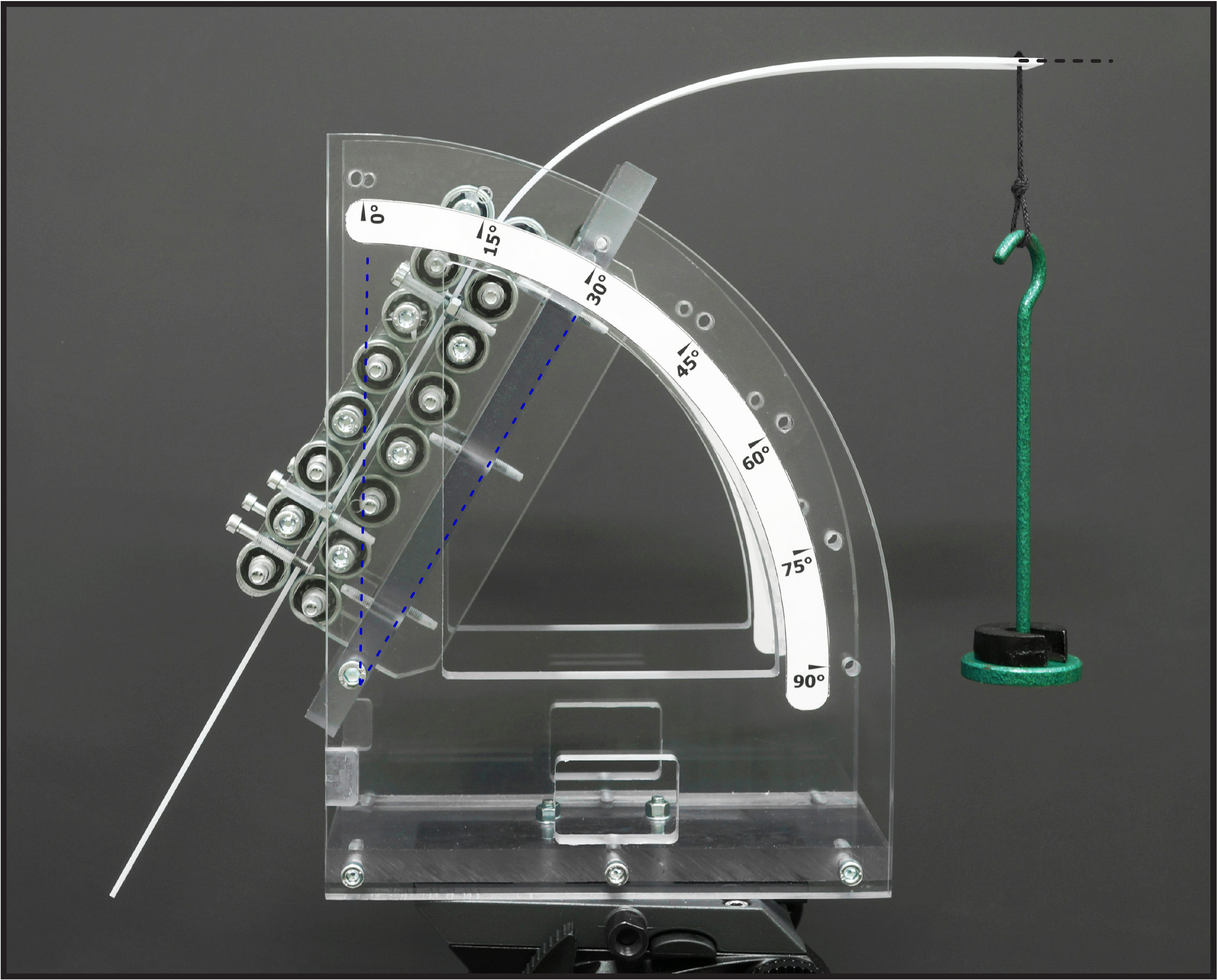}
\caption{The Prototype 1 loaded in a configuration which does not need any counterweight.
The rod used in the prototype is made up of a solid polycarbonate elastic lamina
(inclined at an angle $\alpha$=30$^\circ$ with respect to the vertical direction)
of bending stiffness $B$=0.03 Nm$^2$ and total
length $\bar{l}+ l^*$=0.487 m with one end subject to a dead load $ P_2 $=1.53 N.
The theoretical value of the length defining the equilibrium configuration is $a_{eq}$=0.128 m,
while the value measured on the prototype is equal to 0.126 m.}
\label{fiore3}
\end{center}
\end{figure}
%%%%%%%%%%%%%%%%%%%%%%%%%%%%%%%%%%%%%%%%%%%%%%%%%%%%%%%%%%%%%%%%%%%%%%

As a conclusion, the following modes of use of the elastic scale can be envisaged.

\begin{itemize}
\item The easiest way to use the elastic scale is referring to eq. (\ref{evviva}) and measuring the two angles $\theta_0$ and $\theta_{\bar{l}+l^{*}}$.
Assuming that $P_1$ and $\alpha$ are known, $P_2$ can be evaluated. Note that $B$ is not needed in this mode of use.

\item Another mode of use of the elastic scale is through the measure of the length $a_{eq}$. Knowing $P_1$, $B$, and $\alpha$, $P_2$ can be determined in the following steps:
(i.) eq. (\ref{tratto1})$_1$ gives $\theta_0$;
(ii.) eq. (\ref{evviva}) gives $\theta_{\bar{l}+l^{*}}$ as a function of the unknown $P_2$;
(iii.) eq. (\ref{tratto1})$_2$ provides an equation for the unknown $P_2$, to be numerically solved.

\end{itemize}

Note that eqs. (\ref{tratto1}) define $a_{eq}$ as a one-to-one function respectively of $\theta_0$, eq. (\ref{tratto1})$_1$, and $\theta_{\bar{l}+l^*}$, eq. (\ref{tratto1})$_2$, while eq.
(\ref{evviva}) defines a unique relation between $\theta_0$ and $\theta_{\bar{l}+l^*}$ (within the limits of variability of these two angles).
Therefore, if all the possible deformations of the elastica which are unstable even for clamped end are not considered, the equilibrium solution of eqs. (\ref{tratto1}) and (\ref{evviva}), when it
exists is unique.

\section{Stability} \lb{socmel}

Equilibrium configurations of the proposed mechanical system are expected to be unstable by
observing that a perturbation of the system at an equilibrium position, $a=a_{eq}$, through an increase (or decrease)  of $a$, yields a leftward (or rightward)
unbalanced Eshelby-force, which tends to increase the perturbation itself.
However, instability of the equilibrium configuration exploited in the deformable arm scale does not necessarily represent a drawback, as it could
increase precision in the measure.

In a rigorous way, the instability of a deformed configuration
can be detected by investigating the sign of the second variation of the total potential energy, which can be written as
\begin{equation}
\label{var2ept}
\begin{split}
\delta_\epsilon^{2}\mathcal{V}&=\frac{1}{2}\biggl\{B
\intop_{0}^{a_{eq}}\left[\theta_{var}^{'}(s)\right]^{2}ds+B\intop_{a_{eq}+l^{*}}^{\bar{l}+l^{*}}\left[\theta_{var}^{'}(s)\right]^{2}ds-\sin\alpha\left[P_1\theta^{'}_{eq}(a_{eq})+P_2\theta^{'}_{eq}(a_{eq}+l^{*})\right]a_{var}^{2}\\
&\qquad +P_1\intop_{0}^{a_{eq}}\bigl(\cos\alpha\cos\theta_{eq}(s)-\sin\alpha\sin\theta_{eq}(s)\bigr)\theta_{var}^{2}(s)ds\\
&\qquad
+P_2\intop_{a_{eq}+l^{*}}^{\bar{l}+l^{*}}\bigl(\sin\alpha\sin\theta_{eq}(s)-\cos\alpha\cos\theta_{eq}(s)\bigr)\theta_{var}^{2}(s)ds\biggr\}.
\end{split}
\end{equation}

The second variation, eqn (\ref{var2ept}), becomes
\begin{equation}
\begin{split}
\delta_{\epsilon}^{2}\mathcal{V}&=\frac{1}{2}\Biggl\{B\intop_{0}^{a_{eq}}\left[\theta_{var}^{'}(s)+\frac{\Gamma_1(s)}{B}\theta_{var}(s)\right]^{2}ds+B\intop_{a_{eq}+l^{*}}^{\bar{l}+l^{*}}\left[\theta_{var}^{'}(s)+\frac{\Gamma_2(s)}{B}\theta_{var}(s)\right]^{2}ds \\
&\qquad +\biggl[\Gamma_2(a_{eq}+l^{*})\left[\theta^{'}_{eq}(a_{eq}+l^{*})\right]^2-\Gamma_1(a_{eq})\left[\theta^{'}_{eq}(a_{eq})\right]^2\\
&\qquad -\sin\alpha\left(P_1\theta^{'}_{eq}(a_{eq})+P_2\theta^{'}_{eq}(a_{eq}+l^{*})\right)\biggl]a_{var}^{2}\Biggr\},
\label{eq:varseconda}
\end{split}
\end{equation}
when the two auxiliary functions $\Gamma_1(s)$ and $\Gamma_2(s)$ are introduced as the solutions of the
following boundary value problems
\begin{equation}\label{eq:riccati}
\begin{array}{lll}
\left\{\begin{array}{lll}
\ds\frac{\partial \Gamma_1(s)}{\partial s}+P_1\cos\alpha\cos\theta_{eq}(s)-P_1\sin\alpha\sin\theta_{eq}(s)-\frac{\Gamma_1(s)^{2}}{B}=0,
\qquad s \in [0,a_{eq}],\\[5mm]
\ds \Gamma_1(0)=0
\end{array}
\right.
\\[10mm]
\left\{\begin{array}{lll}
\ds\frac{\partial \Gamma_2(s)}{\partial s}-P_2\cos\alpha\cos\theta_{eq}(s)+P_2\sin\alpha\sin\theta_{eq}(s)-\frac{\Gamma_2(s)^{2}}{B}=0,
\qquad s \in [a_{eq}+l^{*},\bar{l}+l^{*}],\\[5mm]
\ds \Gamma_2(\bar{l}+l^{*})=0.
\end{array}
\right.
\end{array}
\end{equation}
Considering the necessary condition of stability \cite{Majidi}, an equilibrium configuration can be stable if
\begin{equation}
\begin{split}
\Delta&=\Gamma_2(a_{eq}+l^{*})\left[\theta^{'}_{eq}(a_{eq}+l^{*})\right]^2-\Gamma_1(a_{eq})\left[\theta^{'}_{eq}(a_{eq})\right]^2 -\sin\alpha\left(P_1\theta^{'}_{eq}(a_{eq})+P_2\theta^{'}_{eq}(a_{eq}+l^{*})\right)\geq0 .
\label{eq:condizionepuntualestabilit=0000E0}
\end{split}
\end{equation}
Introducing the Jacobi transformation
\begin{equation}\label{jacobitransformation}
\Gamma_j(s)=-B\frac{\Lambda_j^{'}(s)}{\Lambda_j(s)},\qquad j=1,2,
\end{equation}
which leads to the following Jacobi boundary value problems
\begin{equation}\label{eq:riccatijacobielastica}
\begin{array}{lll}
\left\{\begin{array}{lll}
\ds\Lambda_1^{''}(s)+ \frac{P_1}{B}\bigl(\sin\alpha\sin\theta_{eq}(s)-
\cos\alpha\cos\theta_{eq}(s)\bigr)\,\Lambda_1(s)=0,\qquad s \in [0,a_{eq}],\\[3mm] \Lambda_1(0)=1,\\[3mm] \Lambda_1^{'}(0)=0 ,
\end{array}\right.
\\[14mm]
\left\{\begin{array}{lll}
\ds\Lambda_2^{''}(s)+ \frac{P_2}{B}\bigl(\cos\alpha\cos\theta_{eq}(s)-\sin\alpha\sin\theta_{eq}(s)\bigr)\,\Lambda_2(s)=0,
\qquad s \in [a_{eq}+l^{*},\bar{l}+l^{*}],\\[3mm]
\Lambda_2(\bar{l}+l^{*})=1,\\[3mm] \Lambda_2^{'}(\bar{l}+l^{*})=0 ,
\end{array}\right.
\end{array}
\end{equation}
the auxiliary functions $\Gamma_j(s)$ with $j=1,2$ have been numerically evaluated for all configurations considered in the experiments and,
although conjugate points are not present, the unstable character of the configurations follows from $\Delta<0$.

\paragraph{Instability of the solution at small rotations.}

A general proof that all equilibrium configurations are unstable can be easily derived under the assumption of
small rotations. In this case, the equilibrium configuration can be explicitly obtained
as a function of the length $a$ as
\beq
\theta_{eq}(s,a)=
\left\{
\barr{lll}
\ds \frac{P_1 \sin \alpha}{2 B}(s^2-a^2), \qquad \qquad\qquad\qquad\qquad\qquad\qquad\qquad  s \in [0,a]\\[5mm]
\ds \frac{P_2 \sin \alpha}{2 B}\left[(a+l^*)(a-l^*-2\bar{l}\,)+2(\bar{l}+l^{*})s- s^2\right], \qquad s \in [a+l^{*},\bar{l}+l^{*}],
\earr
\right.
\eeq
so that the total potential energy (\ref{EPT}) is evaluated as
\beq\label{vlineare}
\mathcal{V}(a)=-\frac{\sin^2\alpha}{6 B}\left[P_1^2 a^3+P_2^2 (\bar{l}-a)^3\right]-a\cos\alpha(P_1+P_2).
\eeq

The length $a$ at equilibrium $a_{eq}$ can be obtained by imposing the vanishing of the first derivative
of the total potential energy (\ref{vlineare}), while evaluation of its second derivative at equilibrium results in the following expression
\beq
\left.\frac{\partial^{2}\mathcal{V}(a)}{\partial a^2}\right|_{a_{eq}}=-\sin^2 \alpha \left[a_{eq} P_1^2+ (\bar{l}-a_{eq}) P_2^2\right]<0,
\qquad \forall \,\,a_{eq}\in[0;\bar{l}],
\eeq
demonstrating the instability of all the equilibrium configurations for small rotations.

\paragraph{Stable systems.} Finally, it is worth noting that a deformable scale where the equilibrium configuration is stable
can be easily obtained by adding to the proposed system a linear elastic spring of stiffness $k$, located inside
the sliding sleeve (thus restraining the sliding of the elastic rod, see \cite{tarzan}).
In this case, the stabilizing term $k(a-a_0)^2/2$ (in which $a_0$ is the length $a$ in the unloaded configuration)
is added to the elastic potential energy (\ref{EPT}).
Therefore, findings reported in this article pave the way to the realization of stable systems, would instability
prevent the practical realization of an equilibrium configuration,
which is not the case of the scales shown in the Fig. \ref{system} (left) and Fig. \ref{fiore3}, as shown below.

\section{Prototypes of deformable scale and experiments} \lb{vankuelen}

To test the possibility of realizing a deformable scale, two prototypes (called \lq prototype 0' and \lq prototype 1')
have been designed, produced and tested (at the Instabilities Lab of the University of Trento).

In prototype 0 (shown in Fig. \ref{proto}, left, and in the movie available as supplementary electronic material) the sliding sleeve is 296 mm in
length and is made up of 27 roller pairs (each roller is a teflon cylinder
10 mm in diameter and 15 mm in length, containing two roller bearings).
In prototype 1 (shown in Fig. \ref{system}, left, in Fig. \ref{fiore3} and in Fig. \ref{proto}, right) the sliding sleeve, 148 mm in length, is realized
with 8 roller (Press-Fit Straight Type, 20 mm in diameter
and 25 mm in length) pairs from Misumi Europe.
%%%%%%%%%%%%%%%%%%%%%%%%%%%%%%%%%%%%%%%%%%%%%%%%%%%%%%%%%%%%%%%%%%%%%%
\begin{figure}[!htcb]
\begin{center}
\includegraphics[width= 12 cm]{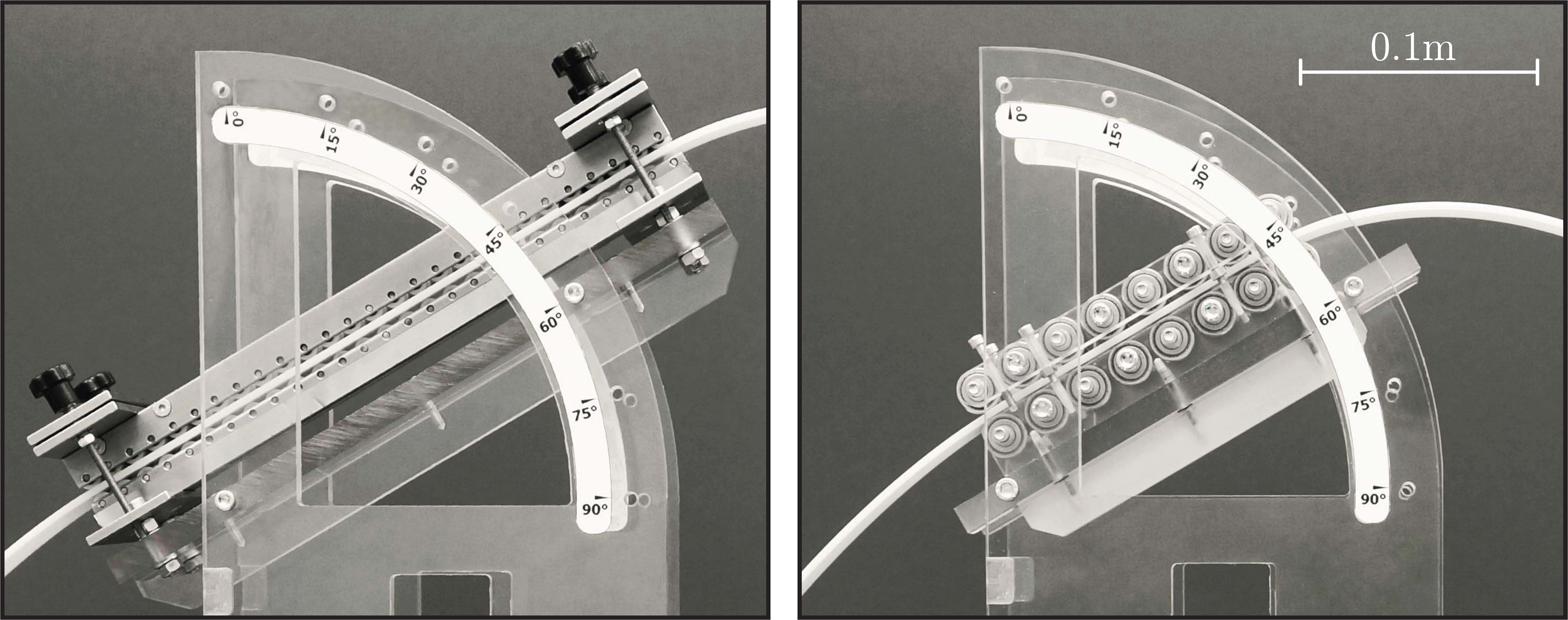}
\caption{Two prototypes of the deformable arm scale: prototype 0 (left) and prototype 1 (right).}
\lb{proto}
\end{center}
\end{figure}
%%%%%%%%%%%%%%%%%%%%%%%%%%%%%%%%%%%%%%%%%%%%%%%%%%%%%%%%%%%%%%%%%%%%%%
The tolerance between the elastic strip and the rollers inside the sliding sleeve can be calibrated with four micrometrical screws.
Two elastic laminas have been realized in solid polycarbonate (white 2099 Makrolon UV from Bayer,
elastic modulus 2250 MPa), one with dimensions 980 mm $\times$ 40.0 mm $\times$ 3.0 mm and the other
487 mm $\times$ 24.5 mm $\times$ 1.9 mm; the latter has been used for the experiments reported in
Fig. \ref{fiore3}, \ref{experiment},  and \ref{sensitivity}, while the former is shown in Fig. \ref{system} (left).
%%%%%%%%%%%%%%%%%%%%%%%%%%%%%%%%%%%%%%%%%%%%%%%%%%%%%%%%%%%%%%%%%%%%%%
\begin{figure}[!htcb]
\begin{center}
\includegraphics[width= 9 cm]{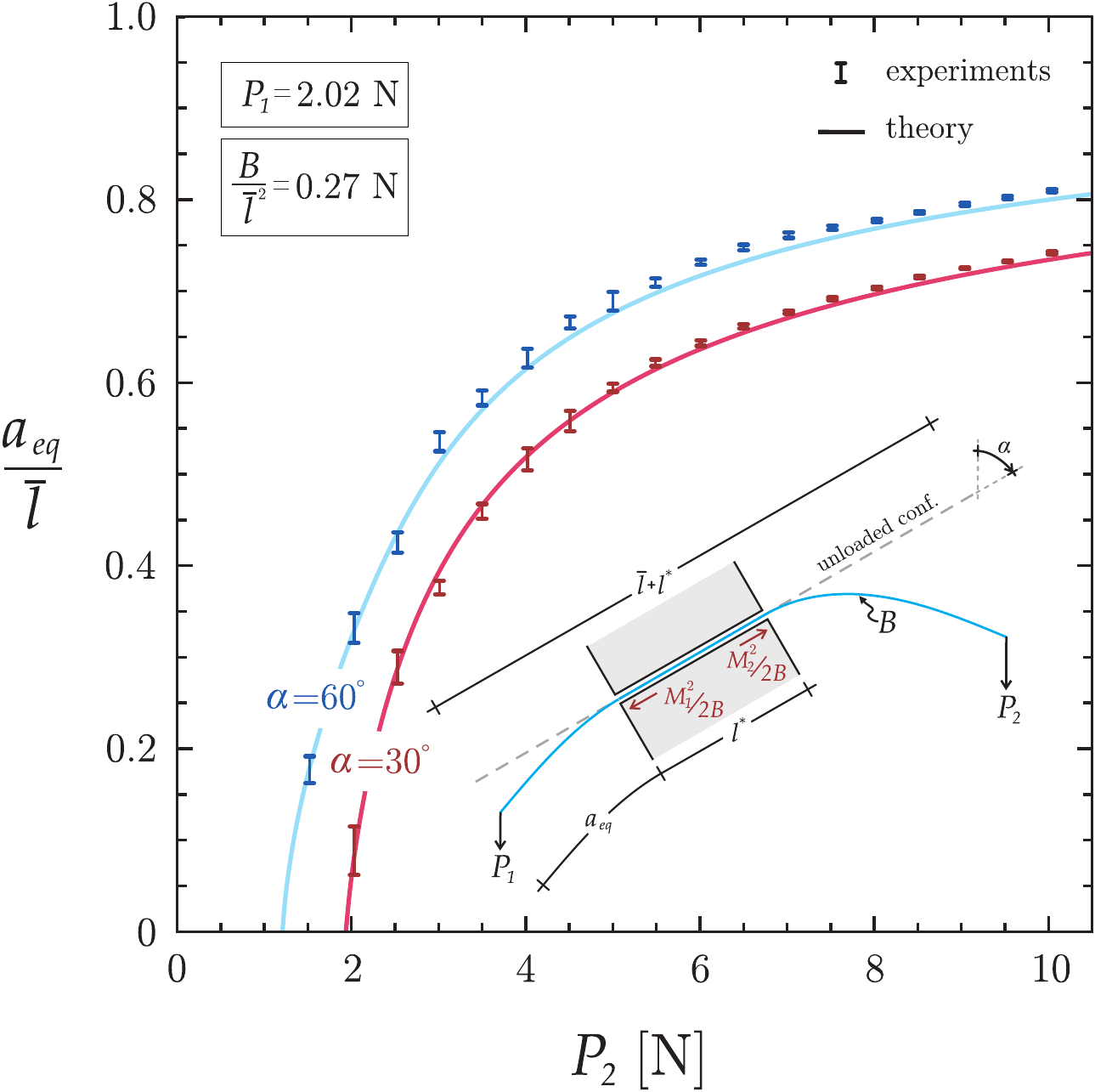}
\caption{Equilibrium length $a_{eq}$ measured on  prototype 1 shown in Fig. \ref{system}
(left, at two inclinations $\alpha$) for different loads $P_2$ versus theoretical predictions.}
\label{experiment}
\end{center}
\end{figure}
%%%%%%%%%%%%%%%%%%%%%%%%%%%%%%%%%%%%%%%%%%%%%%%%%%%%%%%%%%%%%%%%%%%%%%
The sliding sleeve is mounted on a system (realized in PMMA) that may be inclined at different angles $\alpha$.
The two vertical dead loads applied at the edges of the elastic lamina have been imposed manually.
The tests have been performed on an optical table (1HT-NM from Standa) in a controlled temperature $20\pm0.2^{\circ}$
and humidity $48 \pm 0.5\%$ room.

Experimental results (presented in Fig. \ref{experiment} in terms of measured values of the length $a_{eq}$, for different weights $P_2$)
find an excellent agreement with the theory.

The sensitivity of the scale $\mathcal{S}=\partial a_{eq}/\partial P_2$
has been reported in Fig. \ref{sensitivity}, together with the maximum absolute error \lq $err$' found in the experimental
determination of the load $P_2$. The figure
correctly shows that errors decrease at high sensitivity. Moreover, the sensitivity is so high for small $P_2$ that the scale
could in a certain range of use become more accurate than a traditional balance.
%%%%%%%%%%%%%%%%%%%%%%%%%%%%%%%%%%%%%%%%%%%%%%%%%%%%%%%%%%%%%%%%%%%%%%
\begin{figure}[!htcb]
\begin{center}
\includegraphics[width= 11 cm]{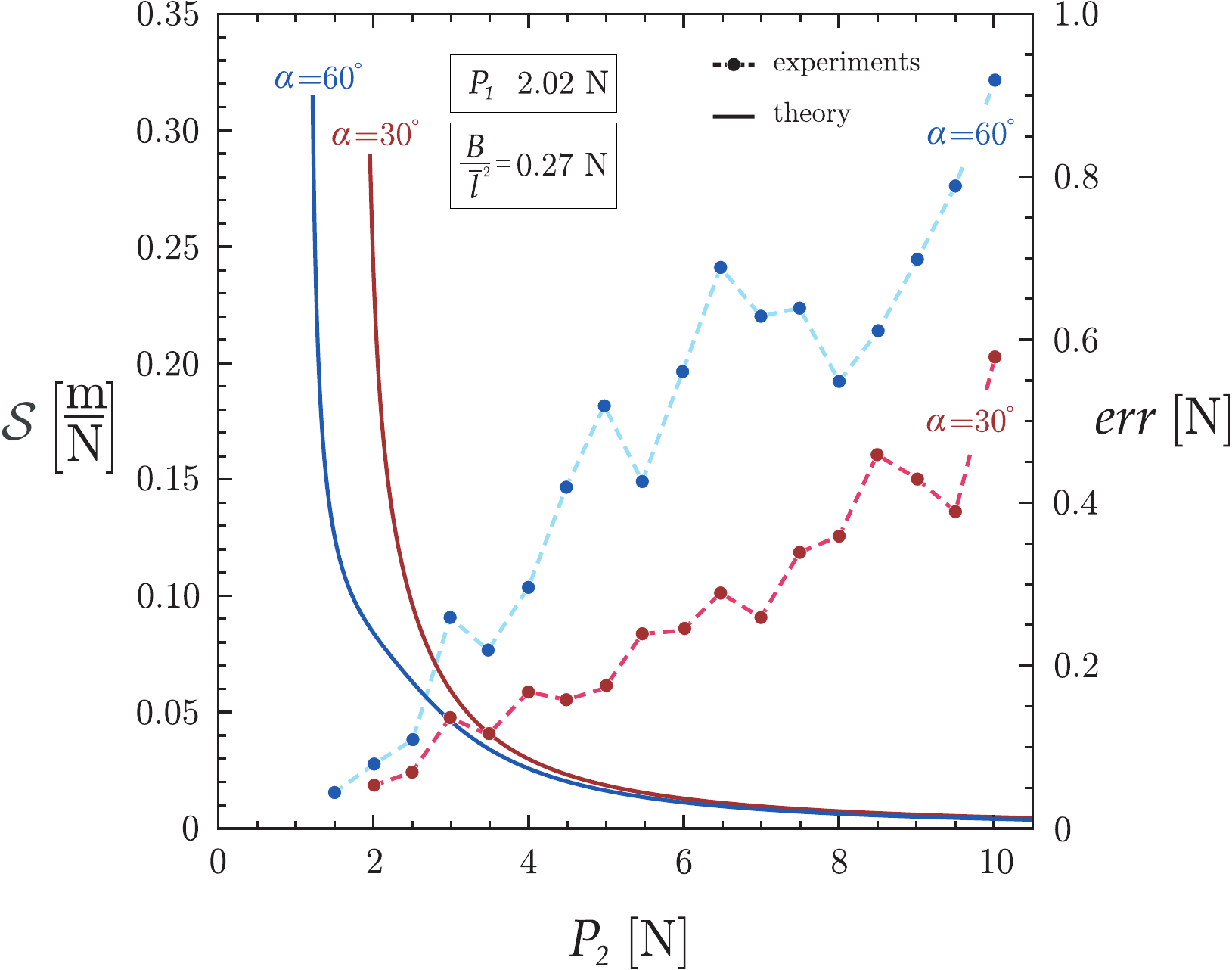}
\caption{Sensitivity (denoted by $\mathcal{S}$) of the deformable arm scale (at two inclinations $\alpha$) as a function of the load $P_2$, reported together
with the maximum absolute error (denoted by $err$) on the loads measured on the prototype shown in Fig. \ref{system} (left).}
\label{sensitivity}
\end{center}
\end{figure}
%%%%%%%%%%%%%%%%%%%%%%%%%%%%%%%%%%%%%%%%%%%%%%%%%%%%%%%%%%%%%%%%%%%%%%

The prototypes represent proof-of-concept devices, demonstrating the feasibility of the elastic scale,
with an accuracy which can be highly improved in a more sophisticated design.
A movie with experiments on the prototypes is available in the electronic supporting material (and at
http://ssmg.unitn.it).

\section{A link to snaking and locomotion}

The
sliding sleeve employed in the realization of the elastic arm scale and considered also in \cite{bigonialtri} and \cite{tarzan} can be viewed
as a perfectly frictionless and tight channel in which an elastic rod can move.\footnote{
The connection with the problem of snake locomotion has been suggested independently by Prof. N. Pugno (University of Trento) and by an anonymous reviewer,
who has correctly pointed out that \lq anyone who's ever tried to use a hand-held snake to unclog a toilet knows very well about this axial reaction force [i.e. the Eshelby-like force]'.}
Our results show that a motion along the channel can be induced even when the applied forces are orthogonal to it.
Moreover, it has been shown that the Eshelby-like forces can have a magnitude comparable with the applied loads.
These forces are the essence of snake and fish locomotion \cite{gray1, gray2, gray3} and must play an important role in the problem of beam snaking occurring during smart drilling of oil wells and in plumbing
\cite{beck}.
Investigation of these and related problems is deferred to another study.

\section{Conclusions}

A new concept for a deformable scale has been introduced in which equilibrium is reached
through nonlinear flexural deformation of the arms and generation of configurational forces. This
equilibrium configuration is unstable,
a feature which can increase the precision of the measure and which does not prevent the
practical realization of prototypes, showing that real balances can be designed and can
be effectively used to measure weights.
The reported findings represent a first step towards applications to deformable systems,
in which nontrivial equilibrium configurations at high flexure can be exploited for
actuators or to realize locomotion.

%%%%%%%%%%%%%%%%%%%%%%%%%%%%%%%%%%%%%%%%%%%%%%%%%%%%%%%%%%%%%%%%%%%%%%%%%%%%%%%%%%%%%%%%

%%%%%%%%%%%%%%%%%%%%%%%%%%%%%%%%%%%%%%%%%%%%%%%%%%%%%%%%%%%%%%%%%%%%%%%%%%%%%%%%%%%%%%%%

%\vspace*{10mm} \noindent
\paragraph{ACKNOWLEDGEMENTS.} The work reported is supported by the ERC Advanced Grant  \lq Instabilities and nonlocal multiscale modelling
of materials' (ERC-2013-ADG-340561-INSTABILITIES).

 { \singlespace
}

\end{document}